# Exploring dimethylsulfate for *in vivo* proteome footprinting

Pasquale Moio[1]*, Arman Kulyyassov[2,3]*, Damien Vertut[2], Luc Camoin,[4,5,6] Erlan Ramankulov[3], Marc Lipinski[2] and Vasily Ogryzko[2§]

[1] Università "G.D'Annunzio", Facolta' di Medicina e Chirurgia, Dip.Scienze Biomediche Sez. Biochimica e Biochimica Clinica, Via dei Vestini,31, 66100 Chieti -Italy

[2] CNRS UMR-8126, Université Paris-Sud 11, Institut de Cancérologie Gustave Roussy, 39, rue Camille-Desmoulins, 94805, Villejuif, France

[3] National Center for Biotechnology of the Republic of Kazakhstan, 43 Valikhanova Str., 010000, Astana, Republic of Kazakhstan

[4] Institut Cochin, Université Paris Descartes, CNRS (UMR 8104), Paris, France

[5] Inserm, U567, Paris, France

[6] Université Paris Descartes, Plate-forme Protéomique Paris 5, Paris, France

*These authors contributed equally in this work.

§ Corresponding author, 33 (1) 42 11 65 25 (phone and FAX), vogryzko@gmail.com

*running title: DMS in 'proteome footprinting'*

**DMS:** dimethylsulfate;

**MALDI-TOF**: Matrix-assisted laser desorption/ionization – Time of flight;

**MRM**: multiple reaction monitoring;

# Summary

**Protein footprinting is a new methodology that is based on probing, typically with the use of mass spectrometry, of reactivity of different aminoacid residues to a modifying reagent. Data thus obtained allow one to make inferences about protein conformations and their intermolecular interactions. Most of the protein footprinting studies so far have been performed on individual proteins *in vitro*. We explore whether a similar approach is possible with the proteins inside of living cells, employing dimethylsulfate (DMS), a reagent widely used for the *in vivo* footprinting of nucleic acids. DMS can induce methylation of the lysine, histidine and glutamate residues on proteins. Using models of the histone H2B/H2AZ heterodimer assembled *in vitro* and from chromatin treated in *vivo*, we show that the methylation by deuterated DMS allows one to distinguish the accessibility of a particular residue in and out of the protein's environmental/structural context. The detection of changes in protein conformations or their interactions *in vivo* can provide a new approach to the identification of proteins involved in various intracellular pathways and help in the search for perspective drug targets and biomarkers of diseases.**

# Introduction

The most common application of mass spectrometry in proteomics is the analysis of native proteomes originating from different biological systems (such as *in vitro* cultured cell, biopsies, biofluids) to identify and quantify proteins and their naturally occurring post-translational modifications [1]. Variations in the amounts of a given protein found in different samples serve as an indicator of their involvement in a pathway under investigation, and thus can aid in its mechanistic analysis as well as in biomarkers research.

However, comparison of different proteomes based only on the analysis of protein levels unnecessarily limits the scope of proteomics methodology. To illustrate this point, one can compare two extreme views of a biological system. A biological system can be considered either 1) as a biochemical reactor, governed by the laws of mass action, or 2) as a mechanical device, where the mutual positions, orientation and interactions between different parts play the most significant role [2]. Whereas, in the first case, the concentrations of different components of the system studied is all that is necessary to understand its functioning, in the second case the amounts of the different parts do not vary, and the most relevant information will come from the analysis of positions, orientations and interactions between different parts. Given that real biological systems are in between these two extreme alternatives, the current focus of proteomics solely on protein amounts might be discarding most of the useful information about the state of the system studied.

Among a number of methodologies aimed to study protein conformation and surface, 'protein footprinting' approach has been developed to probe the surface and/or conformational state of proteins with the help of mass spectrometry [3-8]. The general principle of this approach resides in the induction of chemical modifications of a protein followed by identification by mass spectrometry of the residues that have been modified. Quantitative comparison of the occurrence of the same modification on a protein in different physical states reflects the changing reactivity of the residue and therefore provides useful insights about the interaction surface and/or conformational state of the protein of interest.

Most of the protein footprinting approaches have focused on studying individual proteins *in vitro* [3-7]. Protein footprinting studies on living cells have only begun [8], and have been limited to the analysis of cell-surface proteins, due to their accessibility to the non-permeating reagents typically used in such experiments. We aimed to develop a related approach termed here 'proteome footprinting', which differs from the conventional 'protein footprinting' in two main aspects. First, it is aimed at analysis of the proteins inside living cells, and thus requires the use of membrane-penetrating reagents. Second, it is focused not on a particular protein, but rather on proteome-wide profiling complementary to the proteome-wide analysis of protein expression levels. The rationale is that, instead of protein amounts, the detection of changes in protein surfaces or conformations *in vivo* could also serve as an indication of the protein's involvement in a particular pathway or cellular response, regardless of what these changes mean for the mechanism of the protein's action.

As a proof of principle, we have chosen to explore protein methylation induced by the alkylating agent dimethyl sulfate (DMS). This membrane permeating reagent has been previously used for *in vivo* and *in vitro* footprinting of nucleic acids (DNA and RNA) [9-11]. However, the proteins can also be among the targets of DMS treatment. In our work, we set out to explore several questions about the potential of DMS as a reagent for proteome footprinting *in vivo*. In particular, we confirm that DMS can modify many proteins *in vivo* and found several types of amino acid residues that we repeatedly detect as having been modified by DMS.

We also explored the possibility of using DMS-induced methylation in quantitative studies. Stable isotope labeling, either *in vivo* or *in vitro*, has become an established approach in quantitative proteomics [12, 13]. Here we propose a stable isotope utilizing scheme based on methylation of denatured protein with deuterated DMS (DMSD6). For any particular amino acid residue, a quantitative comparison between the methylation by regular DMS of the protein in context of its tertiary and quaternary structure (e.g., *in vivo*) and the methylation by DMSD6 of the denatured protein (typically in gel after SDS-PAGE separation) allows us to distinguish between the reactivities of this residue in and out of the particular structural context. Thus, the methylation of a denatured protein with DMSD6 establishes the common reference baseline, which can be further used for comparison between the states of this residue, now in different structural (and physiological) contexts.

To show the applicability of this general methodology, we have chosen chromatin as a model for our study. In the last two decades, chromatin has emerged as an important factor in genome function in eukaryotes, including its role as a carrier of epigenetic information [14]. Epigenetic information exists in addition to genetic information; it is carried in molecular structures other than DNA, and cannot be deduced from genomic sequence alone. Many functional states of chromatin have been characterized, differing in localization inside the nucleus, in the degree of folding, in post-translational histone modifications, such as methylation, ubiquitylation, sumoylation, phosphorylation and acetylation, as well as in the presence of various histone variants, such as H3.3, H2AZ, H2ABBD and macro-H2A. The development of approaches to monitor the structural changes of the chromatin proteins *in vivo* will help to further explore the differences between the alternative functional states of chromatin and to elucidate the mechanisms of eukaryotic genome function and the mechanisms of processing and coding of epigenetic information.

## Materials and methods

**Recombinant DNA, cell culture and Western analysis**

To produce recombinant histones (for in gel DMS treatment and H2AZ/H2B dimer reconstitution), we used Luger's plasmids encoding for *X. laevis* histones H3, H4 and H2B [15], whereas the human H2AZ expressing gene was subcloned from the pOZFHH vector [16] into the Pet28c vector. For most of the *in vivo* footprinting experiments, HeLa cells were grown in DMEM with high glucose (PAA) and 10% FBS (PAA). Mitotic HeLa cells were prepared by first arresting cells with 2 mM hydroxyurea for 15 h, which were then released by washing with PBS and fed with fresh DMEM medium for 6 h before the addition of 0.5 μg/ml colcemid and harvesting 2 h later. Histones from HeLa cells were prepared by acid extraction according to a standard protocol [17]. The proteins were separated on 18% Novex Tris-Glycine precast gels, following the manufacturer's protocol, and stained with Coomassie Brilliant Blue for protein visualization. Ethanol was used for gel fixation in all experiments. For the analysis of the histone octamers, the HeLa S3 cells expressing the epitope-tagged histones (e:H2A and e:H2AZ) were generated previously [16]. For the Western analysis, HeLa cells were directly lysed in SDS-sample buffer and the lysates were separated on a 4-12% Novex Tris-Glycine precast gels. Separated proteins were transferred to nitrocellulose membranes and probed with antibodies according to the standard procedure [18]. Antibodies against methylated lysine (ab23366) were from Abcam.

**Treatment with DMS *in vivo* and *in vitro*.**

For the *in vivo* treatment, the cells were washed and again resuspended in PBS (phosphate buffered saline: 137.93 mM NaCl, 2.67 mM KCl, 1.47 mM $KH_2PO_4$, 8.1 mM $Na_2HPO_4$, pH 7.4, Invitrogen), then 1% DMS was added for a 5 min incubation at room temperature. DMS was quenched by addition of 125 mM glycine, and the cells were washed twice with PBS.

For the *in-vitro* treatment of histone dimers, 1 µg of a H2AZ/H2B dimer preparation was diluted in 30 µl of PBS, and treated with 1% DMS at room temperature. The reaction was stopped by adding ammonium hydroxide to 1 M final concentration [19]. The sample was dried on Speedvac, resuspended in LDS loading buffer (Invitrogen), run on 18% Novex Tris-Glycine precast gels, following the manufacturer's protocol, and stained with Coomassie Brilliant Blue for protein visualization.

For the *in-gel* treatment, the gel slice with the target protein was resuspended in PBS and treated for 5 min with 1% DMS at room temperature. Ammonium hydroxide was added to 1 M final concentration to stop the reaction. The gel slice was washed twice with 50 mM ammonium bicarbonate, then processed for the trypsin digestion, as described in the **mass spectrometry** section. To choose the DMS doze used in most of the experiments, titration experiments were performed. Different concentrations and incubation times were used, in order to find a DMS doze that gives a detectable peak that is still significantly smaller than the peak corresponding to the unmodified peptide (Figure S1).

**H2AZ/H2B heterodimer reconstitution**

The *E. coli* cell pellets expressing the *X. leavis* histone H2B and human 6XHis-tagged H2AZ were dissolved in guanidine-containing solubilization buffer (20 mM Tris HCl pH 8, 10% Glycerol, 0.2 mM EDTA, 0.1% Tween, 450 mM KCl and 6 M guanidine hydrochloride), then centrifuged to remove undissolved material. The extracts were mixed at equimolar ratios of recombinant proteins and co-renatured by dialyzing with the same buffer containing 0.1 M guanidine [20]. The protein complex was purified by $Ni_2$-NTA agarose chromatography (Qiagen), using the manufacturer's protocol. The H2AZ/H2B dimer was further purified from the histone monomers by size-exclusion chromatography [21] and stored in PBS buffer (Figure S2).

**Purification of the e:H2AZ and e:H2A-containing octamers from HeLa cells**

The nuclear pellet was prepared from the HeLa cells expressing epitope-tagged versions of the H2A and H2AZ histones, according to the protocol described in reference [22]. The pellet was resuspended in an equal volume of nuclease digestion buffer (0.34 M sucrose, 10 mM Tris [pH 7.5], 3 mM $MgCl_2$, 1 mM $CaCl_2$). The nuclei were incubated with micrococcal nuclease for 20 min at 37°C until the reaction was stopped with 4 mM EDTA. To release the undigested chromatin, the pellet was sonicated and the desired size of chromatin fragments (200-500 bp) was monitored by use of a 2% agarose gel. The unsolubilized material was removed by centrifugation at 16,000 $g$ for 30 min, and the supernatant was dialyzed against 300 mM KCl in 20 mM Tris (pH 8.0), 10% glycerol, 0.1% Tween overnight. The precipitated material was removed by centrifugation at 45,000 $g$ for 30 min. The double-tag purification was performed as described in reference [16]. Briefly, 5 ml of chromatin solution was incubated overnight with 500µl (packed volume) of the anti-FLAG agarose beads (Sigma) and washed five times with 300 mM KCl in 20 mM Tris (pH 8.0), 10% glycerol, 0.1% Tween. The tagged histone-containing chromatin fragments were eluted with 500 µl of FLAG peptide (1 mg/ml FLAG peptide in 300 mM KCl in 20 mM Tris [pH 8.0], 10% glycerol, 0.1% Tween). 50 µl of Ni-NTA-agarose was added to the FLAG eluate and incubated for 2 h. The Ni-NTA beads were washed with the same washing buffer. The final elution with 250 mM imidazole (in 20 mM Tris [pH 8.0], 10% glycerol, 0.1% Tween) was done after the elution with 2 M NaCl in 20 mM Tris (pH 8.0), 10% glycerol, 0.1% Tween.

**Mass spectrometry**

The gel slices were dehydrated with 300 µl of 50% acetonitrile followed by 300 µl of 100% acetonitrile, then re-hydrated with 300 µl of 50 mM ammonium bicarbonate. A final dehydration was performed with 2 washes of 300 µl of 50% acetonitrile, followed by 2 washes of 300 µl of 100% acetonitrile. Each wash was carried out for 10 min at 25°C with shaking at 1400 rpm. The gel slices were dried in a speed-vac at 35°C for 10 min. For trypsin digestion, the gel slices were pre-incubated with 7 µl of 15 ng/µl trypsin (Promega France, Charbonnières) at room temperature for 10 min. Afterwards, 25 µl of 50 mM ammonium bicarbonate was added, and the gel slices were incubated at 37°C for 16 h. The peptide-containing supernatants were dried at 56°C by speed-vac for 30 min, then resuspended in 20 µl of solution containing 0.05% formic acid and 3% acetonitrile for mass spectrometry experiments.

For the MALDI-TOF analysis, the digested samples were spotted directly onto a 96x2 MALDI Plate (Applied Biosystems, Foster City, CA, USA) using the dried-droplet method (0.5 µl of the sample) and mixed on the plate with an equal volume of 5 mg/ml alpha-cyano-4-hydroxycinnamic acid (CHCA) matrix in ACN/water/trifluoroacetic acid (50/50/0.1%) (CHCA Kit, LaserBio Labs, Sophia Antipolis, France). Droplets were allowed to dry at ambient temperature. The analyses were performed on a Voyager DE-PRO mass spectrometer (Applied Biosystems, Framingham, MA) and acquired in positive ions using reflector mode.

Spectra were obtained in a reflectron-delayed extraction in source over a mass range of 500-3500 Daltons. Raw spectra were treated manually using a noise filter algorithm (correlation factor 0.7) and the default advanced baseline correction. Then internal calibration using auto-digestion tryptic peptides or external calibration using a mixture of five external standards (PepMix 1, LaserBio Labs, Sophia Antipolis, France) was performed. The intensities of the signals were measured by calculating the peak areas. Specifically, the data were first inspected on the Data Explorer TM software, and after subtraction of the background, the files were converted to ASCII format, exported to MS Excel and the total ion count corresponding to a specified area was integrated. The L/H ratios (the area L of the light peak versus the area H of the heavy peak) was calculated for every experiment. The the L/H ratio averages and standard deviations for at least three experiments were calculated.

For MRM analysis, the peptides were analyzed with a nano-HPLC (Agilent Technologies 1200) directly coupled to ion trap mass spectrometer (Bruker 6300 series) equipped with a nano-electrospray source. The separation gradient was 7 min from 5% to 90% acetonitrile. The fragmentation voltage was 1.3 V. The analysis of the spectra was performed with the DataAnalysis for 6300 Series Ion Trap LC/MS Version 3.4 software package.

## Results

**1. DMS can methylate lysine, histidine and glutamate residues in proteins.**

To test whether DMS can induce modifications in proteins, we treated recombinant histones, purified from *E.coli*, with various doses of DMS, directly in gel after SDS-PAGE separation. The proteins were then in-gel digested with trypsin and analyzed on MALDI-TOF. Comparison between the DMS-treated and untreated samples showed that some (but not all) peptides acquired several additional peaks (supplementary file), which differed from the original/parental peaks by n*14 amu (where n = 1, 2, and sometimes 3). Two examples are shown in Figure 1A,B (left), for the H2A peptide HLQLAIR and the H4 peptide ISGLIYEETR. The list of peptides that we found to be modified in our experiments is presented in Table 1. To identify which residues have been modified by the DMS treatment, the same samples were analyzed with LC-MS/MS. The mass spectrometer was run in MRM mode, i.e., was set to isolate and CID-fragment the preselected set of ions with particular m/z ratios. The residues identified are presented in Table 1, and supporting MS/MS spectra in the Supplementary Figure S3. As expected, we observed many peptides methylated on the lysine residue. Quite unexpectedly, we also found that histidine and glutamate residues were also targeted often, as shown in Figure 1A,B (right). Strikingly, the H4 peptide ISGLIYEETR has two glutamates that can be modified, and HPLC allowed us to separate two alternatively modified forms of this peptide, each corresponding to a different glutamate residue being affected (Figure 1B, bottom left).

We conclude that, consistent with the fact that methylation of proteins on the histidine and glutamate residues has previously been observed in nature [23, 24], DMS can also induce these modifications. Figure 1C presents a scheme showing how this methylation can occur in accordance with the principles of organic chemistry. The rather broad range of DMS targets makes it an attractive reagent for proteome footprinting analysis, since it can be more informative as compared to specific reagents targeting a single specific residue.

**2. The use of deuterated DMS allows one to distinguish between modifications induced in different environmental contexts in the same experiment.**

To confirm that the source for the methyl groups on the modified amino acid residues was DMS, we used a deuterated version of this reagent – DMSD6, which has the 6 hydrogen atoms replaced by deuterium atoms. As seen in Figure 2A, now the modified peptide differs from the parental peptide by 17 amu, confirming that the methyl groups originated from DMS.

The difference of 3 amu is sufficient to distinguish between the light and heavy versions of the modified peptides on most modern mass spectrometers. Accordingly, we next asked whether combining the treatments with light and heavy versions of DMS in a single experimental sample would allow us to compare the reactivity of a particular residue in the protein placed in different environmental contexts. As a model, we expressed recombinant histone H2A.Z in bacteria and treated the cells by ordinary DMS. Afterwards, the histone was extracted from the cells and methylated again, now with DMSD6. The protein was separated on SDS-PAGE and in-gel trypsinized, then the tryptic spectra were analyzed by MALDI-TOF. As seen in Figure 2B, three modified peptides have been found in this experiment. However, the relative intensities of the modifications induced *in vivo* and *in gel* (L/H ratios, corresponding to shifts of 14 amu and 17 amu, respectively) are different, particularly when the TTSHGR and ATIAGGGVIPHIHK peptides are compared. Our data confirm that this methodology can be used to differentiate between the reactivities of a particular amino acid residue in different conditions, such as the *in vivo* versus *in vitro*.

**3. Folding in the context of H2B/H2AZ dimer induces dramatic changes in residue reactivities compared to the unfolded H2B and H2AZ proteins.**

Since histones are absent in bacteria, the interpretation of the L/H ratios, obtained in the previous experiment, as residue reactivities *in vivo* versus *in vitro*, albeit formally correct, is not physiologically relevant. Moreover, histone overexpression in bacteria causes most of the protein to aggregate and form inclusion bodies [15]. In addition, strong evidence exists that the individual H2A and H2B polypeptides could be folded, stable entities only when they are complexed as the H2A/H2B dimer [25, 26].

To use a more relevant experimental system, we used H2AZ/H2B heterodimer, where both H2AZ and H2B histones are expected to be in their folded state. We tested whether it was possible to find amino acid residues in these histones that react with DMS differently when these proteins are present in the denatured state or as a part of the H2AZ/H2B heterodimer. The H2AZ/H2B heterodimer was reconstituted *in vitro* from recombinant histones expressed in *E.coli* (see the Materials and Methods section and supplementary Figure S2). The dimer was first treated with DMS, next the DMS was inactivated by 1 M ammonium hydroxide, then the proteins were separated on SDS-PAGE and in-gel treated with DMSD6, after which the proteins were in-gel digested with trypsin and analyzed by MALDI-TOF.

Figure 3A and Table 2 show that, for the peptides found to be methylated, the ratios of light to heavy versions of their modified forms range from 0.3 to 4.0, indicating either increasing reactivity or suppression of methylation in the context of the native fold (i.e., H2AZ/H2B heterodimer). In particular, we observed that L/H ratios for the H2AZ peptides vary from 0.3 (for ATIAGGGVIPHIHK peptide) to 1 (for TTSHGR peptide), which is significantly higher than the variation between corresponding L/H values in the experiments with individual histone H2AZ (Figure 2B). These data suggest that, at least for some of the residues, their reactivity strongly depends on whether the histone was a part of the H2AZ/H2B heterodimer or was an individual unfolded molecule. Figure 3B shows the positions of the methylated residues along the amino acid sequences of the H2AZ and H2B histones.

**4. The reactivity of ISGLIYEETR peptide of histone H4 in the context of the H2AZ histone octamer is similar to its reactivity in mitotic chromatin.**

In mitosis, chromatin undergoes structural changes that include tighter packaging and post-translational modifications. Therefore, one might also expect changes in the surfaces and/or conformations of the histones. Given that the ultimate application of 'proteome footprinting' is the *in vivo* comparison of protein surfaces and/or conformations in different physiological states, we decided to use mitotic chromatin as an experimental model to test whether our stable isotope utilizing scheme can be also used in living cells. We DMS-treated HeLa cells under two different physiological conditions: asynchronous cultures and blocked in mitosis. Histones were prepared from the chromatin of these cells, separated on SDS-PAGE, in-gel treated with DMSD6 and then digested with trypsin. For the analysis, we have chosen to compare two peptides that have shown consistent modification and given a good signal in our previous experiments: the ISGLIYEETR peptide from histone H4 (mass 1180) and the HLQLAIR peptide from histone H2A (mass 850). As one can see from Figure 4A, the ISGLIYEETR peptide exhibits a striking difference in reactivities when its methylation status in chromatin from asynchronous cells is compared to that of chromatin from mitotic cells. Whereas DMS can induce higher levels of methylation of this peptide in the asynchronous cells than it can *in gel* (judging by the ratios between the intensities of the light and heavy isotope shifts: 14 amu and 17 amu, respectively) (Figure 4A, top right, 1180 peptide), we could detect very little methylation in the case of mitotic chromatin after a similar treatment with DMS (Figure 4A, bottom right, 1180 peptide). This is despite the fact that chromatin from the mitotic cells is not protected by nuclear envelope, and thus, arguably, is more accessible for DMS. Notably, the HLQLAIR peptide H2A did not show a difference between the asynchronous and mitotic cells in the same experiment (Figure 4A, right part, compare top and bottom, peptide 850).

Previously, using replacement histone H2AZ as a model, we developed a method to analyze the chromatin environment of nucleosomes containing a particular replacement histone, based on affinity purification of the histone octamer, containing the epitope-tagged histone of interest [16]. We therefore decided to compare the *in vivo* DMS-methylation pattern of the H2AZ-containing nucleosome (which constitutes about 5% of total H2A) with that one of a generic H2A-containing nucleosome. The HeLa cells expressing the tagged versions of H2AZ (e:H2AZ) or H2A histones (e:H2A) were treated with regular DMS, and the tagged histone containing octamers were affinity-purified from chromatin. To be able to compare the modification results with the previous data, the histones were separated on SDS-PAGE and in-gel treated with DMSD6, then digested with trypsin, and the same HLQLAIR and ISGLIYEETR peptides were analyzed. The e:H2A nucleosome exhibited a methylation pattern very similar to the pattern of the asynchronous HeLa cells (Compare Figure 4B, top right, with Figure 4A, top right). On the other hand, the H4 histone from the e:H2AZ nucleosome showed a significant protection from methylation, as compared to the e:H2A-nucleosome, whereas the reactivity of H2A histone from the same nucleosome was not dramatically affected. Overall, the methylation pattern of the e:H2AZ nucleosome appeared similar to that one of mitotic chromatin.

**5. Many cellular proteins are methylated after DMS treatment *in vivo*.**

The principal goal of the proteome footprinting technique is to provide a new dimension to quantitative protein profiling, complementary to the comparison of the protein amounts between different proteomes (Figure 5A). In this context, it becomes important to have an estimate of how many proteins are modified in the *in vivo* setting by the footprinting reagent. To qualitatively evaluate the number of proteins methylated by DMS *in vivo*, HeLa cells were treated with different doses of DMS, lysed and the obtained proteome was analyzed by Western using antibodies against methylated lysine (Abcam). As seen on the Figure 5, whereas the untreated cells have very few protein bands with detectable levels of methylation (Fig. 5, lane 1, the two lower bands most likely correspond to histones H3 and H4), a significant number of new bands become detectable after DMS treatment (Fig. 5, lanes 2 and 3). In fact, the pattern of DMS-methylated proteins shows close resemblance to the total proteome profile (Fig. 5B, lane 4, coomassie staining), indicating that most of cellular proteins are modified by DMS, albeit to a different extent.

# Discussion

Depending on one's view of intracellular organization, quantitative information might be not all that it is necessary to know about the state of proteins in the cell. Data on protein-protein interactions and protein conformations, which cannot be accessed by measuring the protein amounts alone, can provide additional relevant knowledge about the physiological state of the biological system under study. Accordingly, by providing a snapshot of the state of protein surfaces and/or their conformations in a particular biological system *in vivo*, the proteome footprinting technique can add a new dimension to quantitative proteomic profiling, useful either for mechanistic studies or else in the spirit of systems biology – i.e., by obtaining and comparing the 'footprinting signatures' and 'footprinting biomarkers' of proteomes in particular conditions. Notably, most of the common proteomic technologies can be employed in this approach.

Many protein-modifying membrane-permeating reagents are commercially available and thus can be used in the proposed approach to proteome footprinting [27]. Here, as a proof of principle, we tested DMS, used previously to modify nucleic acids in both *in vitro* and *in vivo* footprinting experiments. Among its several advantages is its low price and the availability of a stable isotope labeled form, which allows one to develop a quantitative way to compare the differences in residue reactivities. We observed that DMS can modify at least three different amino acids in proteins: lysine, histidine and glutamate. Based on their nucleophilic nature, cysteine, aspartate and arginine could also be targeted, and all these modifications can occur in natural proteins [23, 24]. However, histones contain very few cysteines (namely, Cys96 and Cys110 in histone H3.1), and the tryptic peptide FQSSAVMALQEACEAYLVGLFEDTNLCAIHAK, containing both of these residues, is large and was difficult to observe in our experiments. On the other hand, arginine is abundant in histones, but trypsin does not cleave after methylated arginine [6]; thus it remains possible that this modification may be detectable, but using a protease with a different specificity. Overall, the wide range of amino acid residues affected by DMS suggests that one can use it to probe a large portion of the proteins under study.

We tested the applicability of our method on three systems, all involving histones in different experimental settings. Although clearly a heterological system, when histone H2AZ was expressed in *E. coli*, slight but reproducible differences in the reactivities between different residues upon *in vivo* and *in-vitro* DMS treatments were observed. Although these unexpected findings can hardly bear any biological relevance, they most likely reflect differences in the state of H2AZ molecule after aggregation upon its overexpression in the bacterial cell or interaction with nonspecific targets in bacterial cell (e.g., DNA).

The second system used, H2AZ/H2B heterodimer, is more biologically relevant. Accordingly, as compared to the H2AZ expressed in bacteria, we saw significantly stronger variations in the reactivities between different residues when they were compared in and out of the context of the dimer, consistent with the protein folding during formation of the protein complex. However, one needs to keep in mind two caveats before interpreting these results. First, the notion of 'native fold' of an individual H2A or H2B histone is meaningless out of the context of the H2A/H2B heterodimer. Upon denaturation, the H2A/H2B dimer behaves as a highly cooperative system, melting as a single unit without any detectable intermediates of dissociated, yet folded, H2A and H2B monomers [25, 26]. This suggests that the individual H2A and H2B polypeptides become folded, stable entities only when complexed as the H2A-H2B dimer, and that the major contribution to the stabilization of the structure derives from the coupling between the H2A and H2B dimers. Thus, when interpreting changes in the residue reactivities, it might be impossible to dissociate the contribution of folding of individual H2AZ and H2B molecules from the effects of H2AZ/H2B complex formation.

Second, before interpreting the observed differences in structural/surface terms, one has also to recognize that DMS is a small molecule, able to penetrate membranes and other hydrophobic obstacles. Therefore, although we present the results by mapping the affected residues along the positions of alpha helices involved in the histone hand-shake interaction between the two histones, it might be difficult to interpret the changes in reactivity to DMS as reflecting the changes in steric protection of a particular residue. Generally, interpretation in terms of steric hindrance is warranted when the 'probing molecule' is large, as is the case for DNAse footprinting. However, in the case of DMS-induced methylation, the accessibility of the residue might not be the limiting factor determining its reactivity. Other factors, including the nucleophilic properties of the residue, due to the structure of its electron orbitals, might be more important.

These interpretational challenges suggest that the application of DMS footprinting for structural studies *in vivo* must be complemented by additional methodologies that assess the structure more directly *in vitro*. A 'footprinting signature' for a particular interaction (or a conformation change) can first be obtained *in vitro*, together with independent proof of the interaction by alternative methods. Afterwards, the *in vivo* and *in vitro* signatures can be compared, allowing one to conclude whether the interaction takes place *in vivo*.

Alternatively, and regardless of their structural/surface interpretation, the DMS footprinting signatures of proteomes *in vivo* could be used to compare different functional states of particular biological system in the spirit of systems biology. A change in protein amounts (or a change in phosphorylation levels) in certain condition often indicates the protein's involvement in the cellular response to this condition. Similarly, the changes in the reactivity to DMS treatment might indicate that the conformation or surface of the protein has changed. This might equally point to its involvement in the response, regardless of what do these changes in reactivity mean in structural/surface terms. Certainly, this first step in 'proteome footprinting' analysis does not preclude further work on interpreting the same data in terms of exact structural and/surface changes. But given an immense volume of data anticipated from such an endeavor, it is reasonable to consider these two goals separately, as the first step alone could already be quite informative. The application of this technology will significantly benefit from using a high resolution mass spectrometer, such as those typically employed in the proteome profiling experiments.

Concerning the third experimental system, chromatin from living cell, we have observed a dramatic decrease in DMS reactivity of the peptide ISGLIYEETR of histone H4 in mitotic cells. It is tempting to interpret these changes as being a result of the more compact chromatin structure protecting the protein surface from interaction with DMS. However, given the above considerations, alternative explanations are equally valid and thus warrant further investigation. On the other hand, regardless of the structural interpretation of the observed differences, the value of our result is that it could provide a new signature mark to distinguish between the two chromatin states, paving the way for classifying many other chromatin states in terms of the closeness of their resemblance to either mitotic or interphase chromatin.

Intriguingly, the methylation signature of the nucleosome containing a replacement histone variant H2AZ appears to be similar to that one of mitotic chromatin. The abundant literature on this highly conserved histone is rather controversial, as it has been shown to be associated with both gene repression and activation [28], and also stabilizing or destabilising the nucleosome in different contexts [29, 30]. Although our data could help in elucidating the functional role of this protein, the search for their deeper interpretation is beyond the scope of this work.

Notably, the novel stable isotopes-utilizing scheme, explored in this work, requires for modification levels to be sufficiently low. This is because the intensities of both light and heavy modifications have to be comparable with each other. However, the residues modified in the first round become refractory to further modification, thus negatively affecting the intensity of the second modification. Importantly however, in our case, a large number of peptides exhibited weaker modification by regular DMS, performed first. This indicates that there were sufficient amounts of unmodified peptide remaining for the second round of methylation, by DMSD6.

Finally, concerning the general applicability of the proposed methodology, Western analysis with antibodies against methylated lysines indicates that the number of proteins modified by DMS in HeLa cells could be quite significant, as the pattern of DMS-methylated proteins appears to closely resemble that of the total cellular proteome. On the other hand, we have also noticed that after treatment of a particular protein with DMS followed by the LC-MS/MS analysis of its tryptic digests, many peptides remain unmodified and therefore do not carry any useful information about the state of the protein. Thus, we expect that when applied to the analysis of the total proteome (or a fraction of it), this approach will significantly benefit from an additional enrichment step (Figure 5A). In this regard, whereas antibodies that specifically recognize methylated lysines are commercially available, antibodies for the methylated histidines and glutamates would have to be developed.

To summarize, our method can contribute to the development of new systematic approaches that aim to tap into the potential wealth of information about the dynamic changes in protein conformation and interactions occurring *in vivo*. These approaches have many potentially useful applications in biology and medical science, in particular, in the development of the new concept of a biomarker, encompassing not only increased levels of a protein or its naturally modified states, but also its changed conformation or changed interactions (for example, aggregation) in the diseased cell, which in principle is detectable with the footprinting methodology.

## Acknowledgements

We thank Dr. K. Luger for the plasmids encoding for *X. laevis* histones, and Dr. L. L. Pritchard for critical reading of the manuscript. This work was supported by grants from "La Ligue Contre le Cancer" (9ADO1217/1B1-BIOCE) and the "Institut National du Cancer" (247343/1B1-BIOCE) to VO, and by NCB Kazakhstan (0103_00404) to AK. The MALDI-TOF Voyager De-Pro was obtained by grants from Hoechst-Marion Roussel.

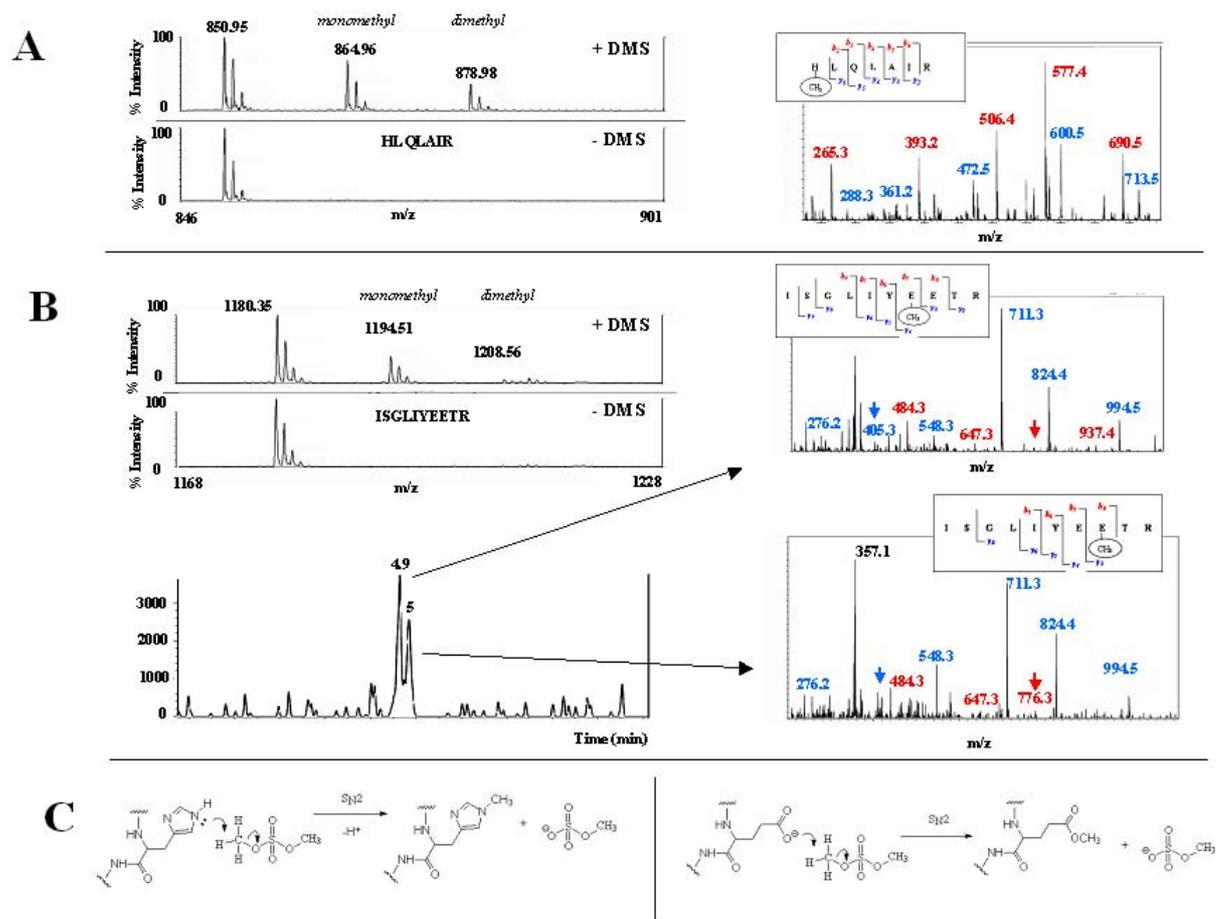

Figure 1

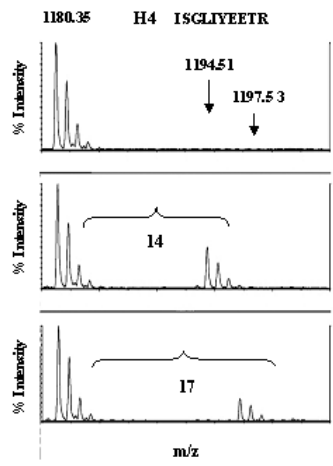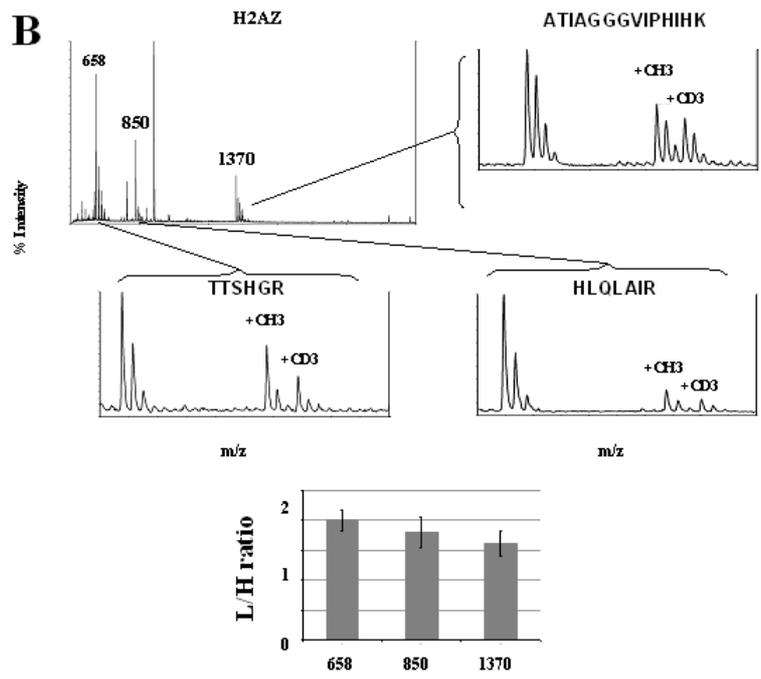

Figure 2

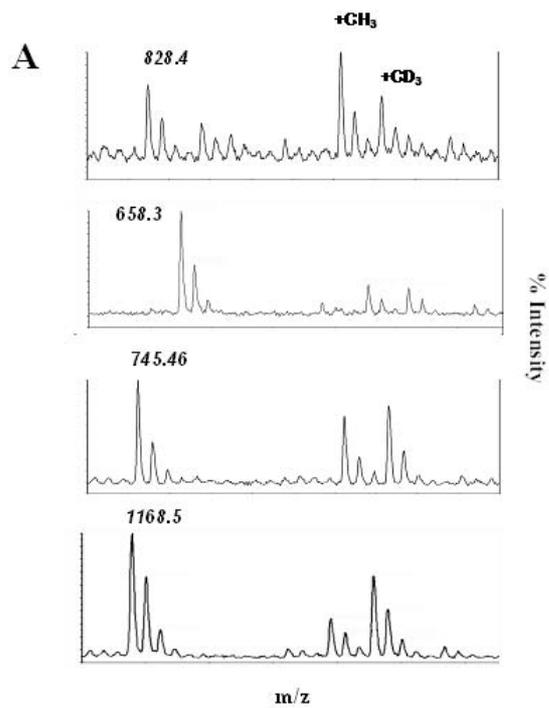
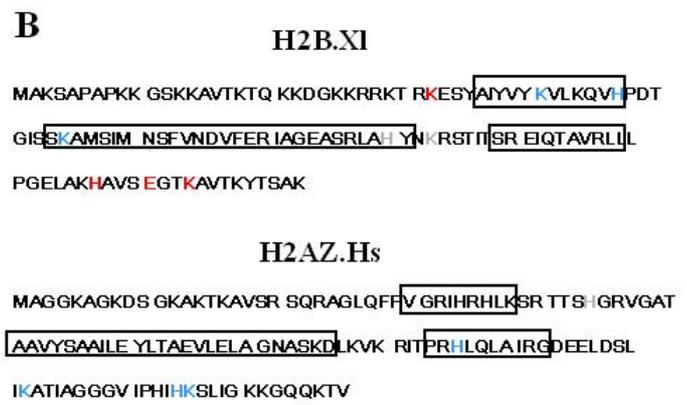

Figure 3

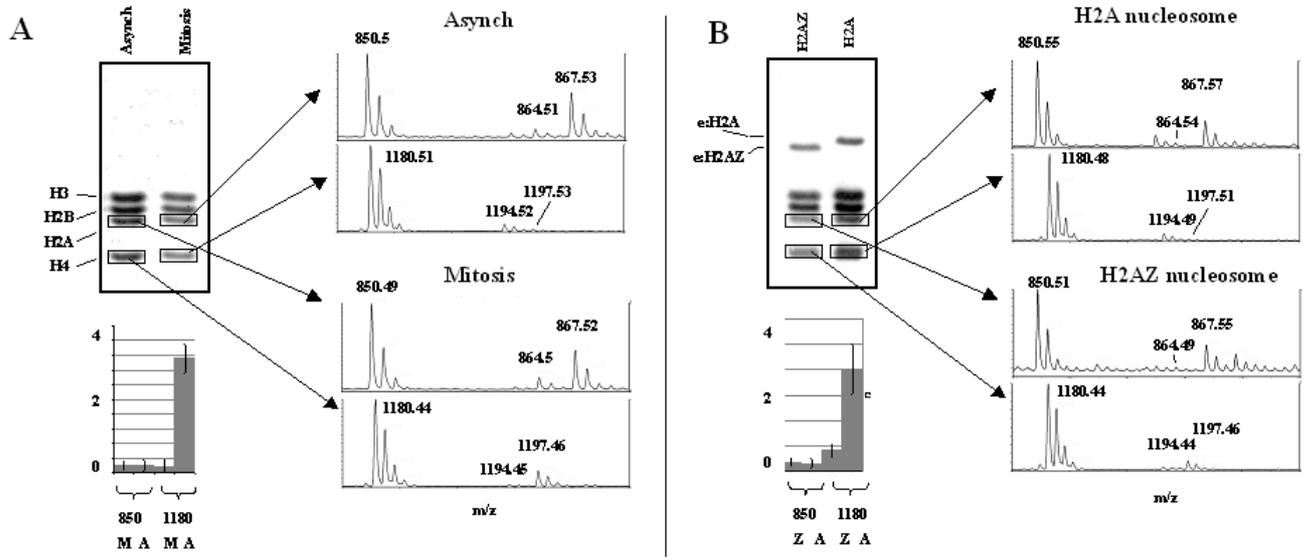

Figure 4

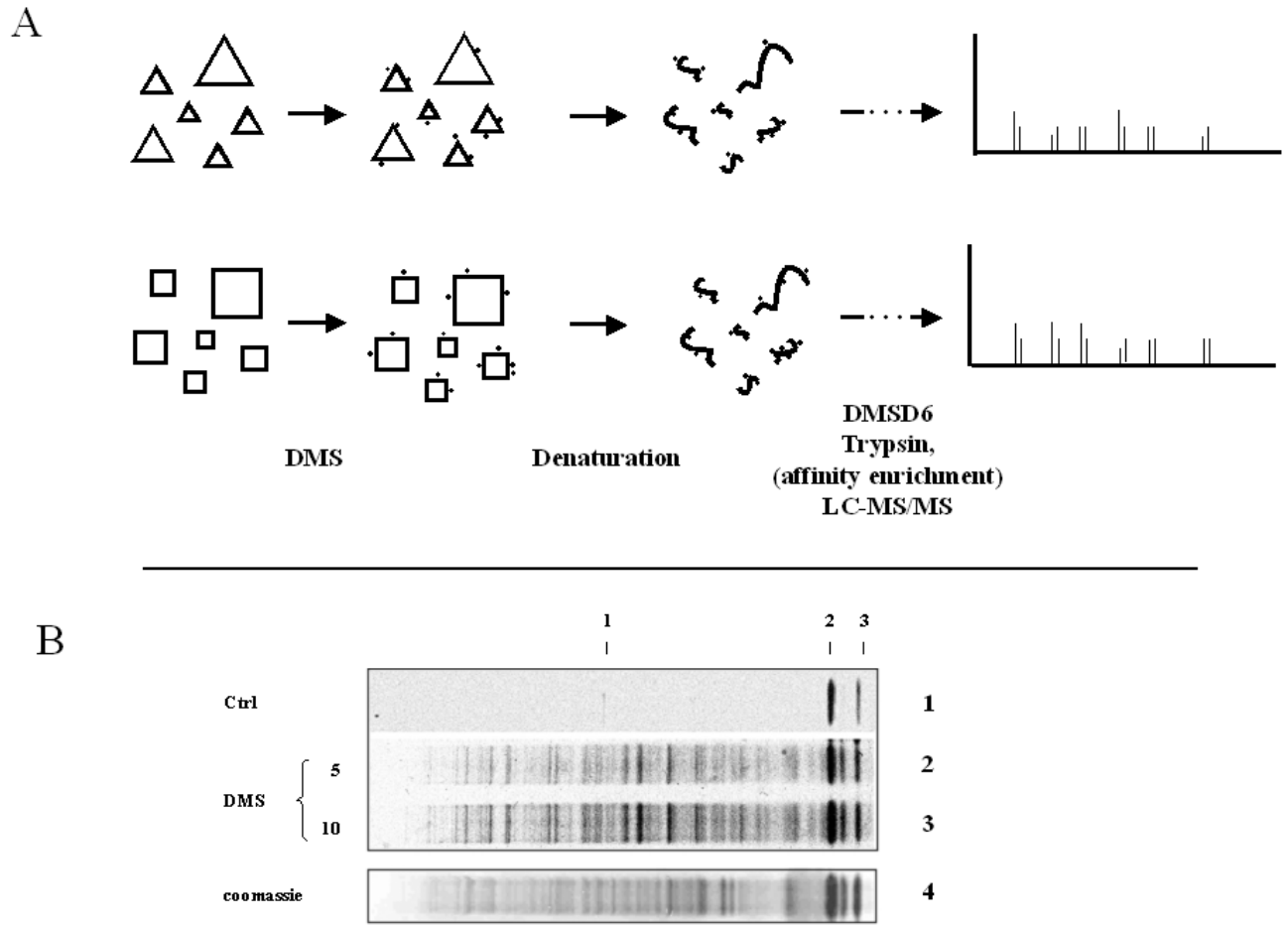

**Figure 5**

# Figure legends

**Figure 1. DMS can methylate lysine, histidine and glutamate residues in proteins.**

**A.** Peptide HLQLAIR from H2AZ histone.

Left: zoomed part of the MALDI-TOF spectrum showing the relevant parental (850) and modified (864) peptides.

Bottom, untreated sample; top, treated sample: abscissa, m/z; ordinate, intensity.

Right: MS/MS spectrum of the 864 ion, showing that His residue is methylated. The y- and b- series ions matching the theoretical spectrum are indicated.

**B.** Peptide ISGLIYEETR from histone H4.

Left MALDI-TOF spectrum showing the parental (1180) and methylated (1194) peptides.

Bottom: HPLC chromatogram from the LC-MS/MS experiment separating the two forms of the modified peptide.

Right: MS/MS spectra of the two forms of methylated peptide. The y- and b- series ions matching the theoretical spectra are indicated. The arrows show the ions with m/z of 405.3 (y3) and 790.3 (b7) that indicate the position of the methyl group in the first peptide (top) and m/z of 420.3 (y3) and 776.3 (b7) that indicate the position of the methyl group in the second peptide (bottom)

C. The direct $S_N2$ displacement mechanism for methylation by DMS is concerted, without an intermediate, and proceeds through a single rate-determining transition state. The DMS molecule is attacked by a nucleophile (nitrogen atom of histidine or lysine residues, or else oxygen atom of carboxylate anion of glutamic acid residue) from the side opposite the leaving group, with bond making occurring simultaneously with bond breaking between the carbon atom and the leaving group.

**Figure 2. The use of deuterated DMS allows to distinguish between modifications induced *in vivo* and *in vitro*.**

A. The ISGLIYEETR from histone H4 treated with DMS or DMSD6. Top, control untreated sample; middle, DMS treated sample; bottom, DMSD6 treated sample. The positions of the light and heavy versions of the methylated peptides are indicated by arrows.

B. H2AZ histone methylated *in vivo* by regular DMS, followed by *in vitro* teatment with DMSD6. Three peptides are shown. Top left, the full MALDI-TOF spectrum; middle and top right, zoomed MS spectra showing that both 14 and 17 forms of methylated peptides are present; bottom, relative intensity ratios of 14/17 methylated forms calculated for different peptides.

**Figure 3. Folding in the context of H2B/H2AZ dimer induces dramatic changes in residue reactivities compared to the unfolded H2B and H2AZ proteins.**

A. Example of differences in methylation between the unfolded histone monomers and H2AZ/H2B dimer. MALDI-TOF spectra for several modified peptides are presented. Abscissa, m/z values; ordinate, signal intensity. The positions of peptides modified by light ($CH_3$) and heavy ($CD_3$) isotope versions of DMS are indicated.

B. Position of the affected residues along the primary amino acid sequence of the H2AZ and H2B histones. The positions of alpha helices involved in the histone hand-shake interaction between the two histones are shown by boxes. The locations of the aminoacids affected are shown in red (in the case of increased reactivity in the context of the dimer) or blue (in the case of decreased reactivity in the dimer context).

**Figure 4. The reactivity of ISGLIYEETR peptide of histone H4 in the context of the H2AZ histone octamer is similar to its reactivity in mitotic chromatin.**

**A:** Mitotic chromatin versus chromatin from asynchronous cells.

Top left. Histones prepared from asynchronous (right) and mitotic (left) cells. SDS-PAGE gel stained with Coomassie Brillian Blue. The bands corresponding to the histones H3, H2B, H2A and H4 are indicated.

Right: Methylation patterns of the H2A HLQLAIR and the H4 ISGLYEETR peptides. The positions of the parental ions (850 and 1180, respectively) and the light and heavy modified versions (864 vs 867 and 1194 vs 1197, respectively) are indicated. Top, asynchronous cells; bottom, mitotic cells.

Bottom left: Quantification of the differences in 14:17 ratios between mitotic (M) and asynchronous (A) cells.

**B:** e:H2A-containing octamer versus e:H2AZ containing octamer.

Top left. Histones prepared from the e:H2AZ (left) and e:H2A (right) containing octamers. SDS-PAGE gel stained with Coomassie Brillian Blue. The bands corresponding to the histones H3, H2B, H2A and H4, and the epitope-tagged e:H2A and e:H2AZ are indicated.

Right: Methylation patterns of the H2A HLQLAIR and the H4 ISGLYEETR peptides. The positions of the parental ions (850 and 1180, respectively) and the light and heavy modified versions (864 vs 867 and 1194 vs 1197, respectively) are indicated. Top, e:H2A octamer; bottom, e:H2AZ octamer.

Bottom left: Quantification of the differences in 14:17 ratios between e:H2A (A) and e:H2AZ (Z) containing octamers.

**Figure 5. The principle of global proteome profiling by in vivo footprinting**

**A:** The general scheme of DMSD6 utilization

Cells in different physiological conditions (top and bottom, respectively) are treated with regular DMS. The cells are lysed, the proteins are denatured and treated with DMSD6, now out of any physiological (and structural) context. After trypsinization and shotgun LC-MS/MS analysis, the *in vivo* reactivities of peptides in several physiological conditions are compared between each other via measuring the ratios between the light and heavy counterparts of their modified versions. Given that the DMSD6 treatment is done with the proteins prepared in identical denatured states, the intensity of the heavy peaks should be identical between the samples and serve as a reference baseline for comparison between the light peaks from different samples. Previous to the LC-MS/MS, affinity purification can be optionally used to enrich the samples in the modified peptides.

**B:** Many cellular proteins are methylated after DMS treatment *in vivo*

1-3: Western analysis with anti methyl-lysine antibodies of the total lysate from HeLa cells. 1 – untreated cells; 2,3 – cells treated with 1% DMS for 5' and 10', respectively. The three endogenously methylated bands are indicated on the top.

4: The same HeLa cells lysate, run in parallel on SDS-PAGE gel, and stained with Coomassie Brilliant Blue.

# Table 1

| N | Histone | m/z MS2 | Sequence, (methylation site) |
|---|---------|---------|------------------------------|
| 1 | H2AZ | 1370.8 | ATIAGGGVIPHIHK(Me) |
| 2 | H2AZ | 1370.8 | ATIAGGGVIPHIH(Me)K |
| 3 | H2AZ | 1118.6 | GDEELDSLIK(Me) |
| 4 | H2AZ | 850.5 | H(Me)LQLAIR |
| 5 | H2AZ | 658.3 | TTSHGR |
| 6 | H2B | 1263.6 | K(Me)ESYAIYVYK |
| 7 | H2B | 1135.6 | ESYAIYVYK(Me) |
| 8 | H2B | 745.4 | LAH(Me)YNK |
| 9 | H2B | 1168.6 | QVHPDTGISSK(Me) |
| 10 | H2B | 1168.6 | QVH(Me)PDTGISSK |
| 11 | H2B | 901.5 | LAH(Me)YNKR |
| 12 | H2B | 901.5 | LAHYNK(Me)R |
| 13 | H2B | 828.4 | HAVSEGTK |
| 14 | H3 | 959.6 | STE(Me)LLIR |
| 15 | H4 | 1134.5 | DAVTYTEH(Me)AK |
| 16 | H4 | 1180.6 | ISGLIYE(Me)ETR |
| 17 | H4 | 1180.6 | ISGLIYEE(Me)TR |
| 18 | H4 | 989.4 | VFLENVIR |
| 19 | H4 | 1325.5 | DNIQGITKPAIR |

**Table 2**

|   | *Sequence (position)* | *Mass* | *Ratio 14/17 ÷ SD* |
|---|---|---|---|
| **H2B** | | | |
| **1** | **K(Me)ESYAIYVYK** | **1263.7** | **430 ÷ 80%** |
| **2** | **ESYAIYVYK(Me)** | **1135.6** | **45 ÷ 20%** |
| **3** | **QVH(Me)PDTGISSK(Me)** | **1168.6** | **48 ÷ 23%** |
| **4** | **LAH(Me)YNKR(Me)** | **901.5** | **120 ÷ 34%** |
| **5** | **LAH(Me)YNK(Me)** | **745.4** | **84 ÷ 12%** |
| **6** | **H(?)AVSE(?)GTK(?)** | **828.4** | **220 ÷ 57 %** |
| | | | |
| **H2AZ** | | | |
| **1** | **TTSH(?)GR** | **658.3** | **112 ÷ 14%** |
| **2** | **H(Me)LQLAIR** | **850.5** | **56 ÷ 9%** |
| **3** | **GDEELDSLIK(Me)** | **1118.5** | **83 ÷ 11%** |
| **4** | **ATIAGGGVIPHIH(Me)K(Me)** | **1370.8** | **29 ÷ 16%** |